\documentclass[letterpaper, 10 pt, conference]{ieeeconf} 

\IEEEoverridecommandlockouts 

\usepackage[english]{babel} 
\usepackage[utf8]{inputenc} 
\usepackage[cmex10]{amsmath} 
\usepackage{amssymb} 

\usepackage{float} 
\usepackage[pdftex]{graphicx} 
\usepackage{amsfonts} 
\usepackage{mathtools} 
\usepackage{xcolor} 
\usepackage{color} 
\usepackage{verbatim} 
\usepackage{graphicx} 
\usepackage{subfig} 
\usepackage{tcolorbox} 
\usepackage{soul} 

{} 
{} 
{} 
\newtheorem{theorem}{Theorem}{} 
\newtheorem{remark}{Remark}{} 
\newtheorem{lemma}{Lemma}{} 
\newcommand{\figref}[1]{\figurename~\ref{#1}} 

\title{\LARGE \bf 
Generalized chi-squared detector for LTI systems with non-Gaussian noise}

\author{Navid Hashemi$^{1}$ and Justin Ruths$^{1}$ 
\thanks{$^{1}$These authors are with the Departments of Mechanical and Systems Engineering at the University of Texas at Dallas, Richardson, Texas, USA {\tt\small navid.hashemi@utdallas.edu, jruths@utdallas.edu}}%
}

\begin{document}

 \maketitle 
 \thispagestyle{empty} 
\pagestyle{empty}

\begin{abstract} 
Previously, we derived exact relationships between the properties of a linear time-invariant control system and properties of an anomaly detector that quantified the impact an attacker can have on the system if that attacker aims to remain stealthy to the detector. A necessary first step in this process is to be able to precisely tune the detector to a desired level of performance (false alarm rate) under normal operation, typically through the selection of a threshold parameter. To-date efforts have only considered Gaussian noises. Here we generalize the approach to tune a chi-squared anomaly detector for noises with non-Gaussian distributions. Our method leverages a Gaussian Mixture Model to represent the arbitrary noise distributions, which preserves analytic tractability and provides an informative interpretation in terms of a collection of chi-squared detectors and multiple Gaussian disturbances. 

\end{abstract}

\section{INTRODUCTION} 

Model-based anomaly detection uses a predictor to forecast the evolution of a dynamical system. This prediction is compared with the actual measured value to test for an appreciable discrepancy, which may indicate the presence of anomaly in the system. If the predictor is perfect, then this task is easy - any discrepancy is enough to cause concern. The task is made more challenging because the predictor has some uncertainty in its forecast, which implies that relatively small discrepancies may be due to this uncertainty rather than anomalous behavior. One of the most fundamental sources of uncertainty in conventional control systems is captured in the terms that quantify system and measurement noise. The overwhelming majority of work on stochastic model-based detection assumes system and measurement noises that are Gaussian distributed, see e.g., \cite{Guo2016,Mo_3,Gupta2}. In the context of linear time-invariant systems especially, normal distributions preserve the capability of finding analytic solutions. In this work, we do not assume any structure of the noise probability density functions; however, we manage to retain some analytic tractability by employing a Gaussian Mixture Model (GMM) representation of these arbitrary distributions. 

In a stochastic context, detectors necessarily trade off sensitivity (rate of true positives - true alarms) for fall-out (rate of false positives - false alarms). The machinery we present to tune detectors is a necessary step to (a) understand the shape of the Receiver Operating Characteristic (ROC) curve, (b) quantify how that shape depends on system and design parameters, and (c) select a point on the ROC that has a desired level of performance - as measured by a tolerable true/false positive balance. In past work we have provided methods to tune several popular detectors when the uncertainty is driven by Gaussian system and measurement noises \cite{Carlos_Justin1,Carlos_Justin2}. When noise distributions deviate significantly from this assumption, then new tools are required to produce an effective means of tuning detectors. Here we provide a generalized chi-squared detector that can be tuned to any desired rate of false alarms through the appropriate selection of the detector sensitivity threshold. 

The process of tuning model-based detectors is to propagate the uncertainty, here the noise distributions, through the system and into the detector. By quantifying the distributions the detector expects to see under normal operation (with no anomalies), we can quantify the trade-off between the true and false positive rates based on the sensitivity threshold we select. We begin by reviewing the tools for the Gaussian case and conclude by demonstrating the backwards compatibility of the results we develop with the Gaussian case and by showing an example. 

\section{MODEL-BASED ANOMALY DETECTION} \label{math} 

Consider a general discrete-time linear time-invariant (LTI) system, 
\begin{equation} \label{eq:LTI} 
\begin{aligned} 
x_{k+1} &= Fx_k + Gu_{k}+v_k\\ 
y_k &= Cx_k + \eta_k 
\end{aligned} 
\end{equation} 
with time index $k \in \mathbb{N}$, state $x_k \in \mathbb{R}^n$, output $y_k \in \mathbb{R}^p$, input $u_k \in \mathbb{R}^m$, matrices $F$, $G$, and $C$ of appropriate dimensions, and iid multivariate noises $v_k \in \mathbb{R}^n$ and $\eta_k \in \mathbb{R}^p$ with covariance matrices $R_1 \in \mathbb{R}^{n \times n}$, $ R_1 \geq 0$ and $R_2 \in \mathbb{R}^{p \times p} $, $R_2 \geq 0$ respectively. The random processes $v_k$ and $\eta_k$ are mutually independent. We assume that $(F,G)$ is stabilizable and $(F,C)$ is detectable such that there are no unstable unobservable or uncontrollable modes. 

The main idea behind fault detection theory is to use an estimator to forecast the evolution of the system in the absence of fault \cite{Patton_1}. In this analysis we consider a model-based estimator of Luenberger form, 
\begin{equation} \label{eq:estimator} 
\hat{x}_{k+1} = F\hat{x}_k + Gu_k + L(y_k -C\hat{x}_k), 
\end{equation} 
where the estimator has perfect model knowledge, $L$ is the observer gain matrix, and $\hat{x}_{k+1}$ is the predicted state. This prediction is compared to the actual measurement received from the sensors. If the difference between the measurement and the predicted output is large, there may be an anomaly in the system. This difference is typically called the residual, 
\begin{equation} \label{eq:residual_def} 
r_k=y_k-C\hat{x}_k, 
\end{equation} 
where $y_k$ is the actual measurement and $C\hat{x}_k$ is the estimated output. Different model-based detectors use the residual in different ways to quantify deviation away from the estimate. One of the simplest and most widely used approach constructs the distance measure $z_k$ from a quadratic form of the residual, 
\begin{equation} \label{eq:chisquared} 
z_k=(r_k-\mu)^T\Sigma^{-1}(r_k-\mu),\qquad z_k>\alpha \longrightarrow \text{alarm} 
\end{equation} 
where $\mu\in\mathbb{R}^p$ and $\Sigma\in\mathbb{R}^{p\times p}$ are the mean and covariance of the residual random vector, $r_k$, under normal operation (no faults or attacks). This detector raises alarms if $z_k$ exceeds an assigned threshold, i.e., $z_k>\alpha$, $\alpha \in \mathbb{R}_{>0}$. 

When the system and sensor noises are zero-mean and Gaussian distributed, then the residual is also a zero-mean Gaussian random variable, i.e., $r_k \sim \mathcal{N}(0,\Sigma)$, with covariance $\Sigma=CPC^T+R_2$, where $P$ represents the asymptotic covariance of the estimation error, $\lim_{k \to \infty} P_k = \lim_{k \to \infty} E[e_ke_k^T] = P$, $e_k=x_k-\hat{x}_k$, and is the solution of the Lyapunov equation \cite{Astrom}, 
\begin{equation}\label{eq:recattiP} 
(F-LC)P(F-LC)^T-P+R_1+LR_2L^T=0. 
\end{equation} 
The distance measure $z_k$ is then a sum of squared Gaussian variables making it chi-squared distributed. For this reason, the detector \eqref{eq:chisquared} is conventionally called the \textit{Chi-Squared Detector}. This name underscores the prevailing assumption that system and sensor noises are Gaussian distributed - the simplifying assumption we lift in this work. 

In past work, we characterized the relationship to select the threshold $\alpha$ in \eqref{eq:chisquared} to yield a desired false alarm rate. Effectively capturing the receiver operator curve (ROC) of the detector, this characterization is critical for comparisons across detectors because all detectors must be tuned for comparable performance. The following Lemma states the relationship for a chi-squared detector with zero-mean Gaussian system and measurement noises. 

\begin{lemma}[\cite{Carlos_Justin2}] \label{lem:chisquared_tuning} 
Assume that there are no anomalies present in an LTI system \eqref{eq:LTI} driven by zero-mean Gaussian system and measurement noises such that the residual $r_k \sim \mathcal{N}(0,\Sigma)$ and consider the chi-squared detector \eqref{eq:chisquared} with threshold $\alpha \in \mathbb{R}_{>0}$. To achieve a desired false alarm rate $\mathcal{A}^*$ set $\alpha = \alpha^{*}:=2P^{-1}(1-\mathcal A^{*}, \frac{p}{2})$, where $P^{-1}(\cdot,\cdot)$ denotes the inverse regularized lower incomplete gamma function. 
\end{lemma} 

\section{NON-GAUSSIAN NOISES}\label{residue} 
Although assuming noises fall according to Gaussian distributions is standard for the theoretical treatment of control systems, practical implementations for tuning chi-squared detectors on real systems can fall short when using Lemma \ref{lem:chisquared_tuning} directly. System noise is typically used to aggregate model uncertainty. For many control concerns, increasing the covariance of Gaussian noise is an effective, conservative approach for capturing model imperfections. However, when we aim to use this noise to predict the expected behavior of the system it is important that the noise distribution is an accurate representation of the model discrepancy. In addition, injecting conservatism into the system noise directly reduces the sensitivity of the detector. For sensor noise, many sensors exhibit nonlinear behavior in the distribution of their reported values or evidence strong quantization effects; both of which can have a dramatic effect on the accuracy of Lemma \ref{lem:chisquared_tuning}. 

\begin{figure}[t] 
\centering 
\includegraphics[width=0.4\textwidth]{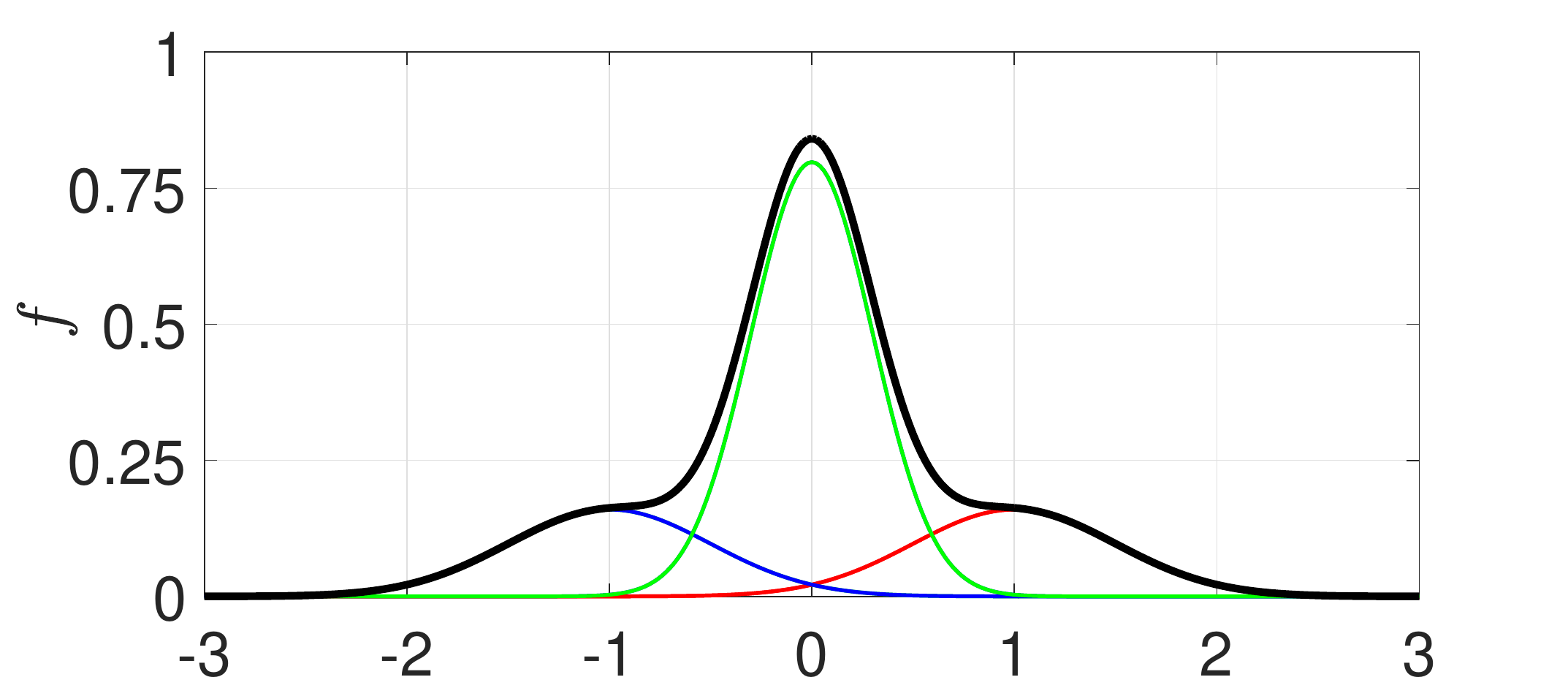}
\caption{Given the probability density function (black) a Gaussian Mixture Model (GMM) expression can be constructed. Here the GMM is composed of three Gaussian modes.} \label{fig:GMM2}
\end{figure} 

We revisit the chi-squared detector defined in \eqref{eq:chisquared} with system and sensor noises that are arbitrary distributed. These distributions can be expressed as convex mixtures of Gaussian distributions \cite{athanassiag.bacharoglou2010}. We construct the so-called \textit{Gaussian mixture model} (GMM) expansion of the distributions for system noise ($v_k$) and measurement noise ($\eta_k$), respectively, 
\begin{align} 
f_{\eta}&=\sum_{j=1}^{m_1} p^{\eta}_j\ \mathcal{N}(x\, |\, \mu^{\eta}_j, K^{\eta}_j), \label{eq:noiseGMM_eta} \\ 
f_{v}&=\sum_{j=1}^{m_2} p^v_j\ \mathcal{N}(x\, |\, \mu^v_j, K^v_j), \label{eq:noiseGMM_v} 
\end{align} 
where $m_1$ is the number of Gaussian modes for the measurement noise and $m_2$ is the number of Gaussian modes for the system noise, $\mu^{\eta}_j \in \mathbb{R}^p$ and $\mu^v_j \in \mathbb{R}^n$ are the mean value of each Gaussian mode, $K^{\eta}_j \in \mathbb{R}^{p \times p}$ and $K^v_j \in \mathbb{R}^{n \times n}$ are their corresponding covariances. It is important that the scalar values $p^{\eta}_j$ and $p^v_j$ satisfy 
\begin{equation} 
\sum_{j=1}^{m_1}p^{\eta}_j=1, \qquad\text{and}\qquad \sum_{j=1}^{m_2}p^v_j=1, 
\end{equation} 
which follows from interpreting \eqref{eq:noiseGMM_eta} (resp., \eqref{eq:noiseGMM_v}) as a total law of probability between Gaussian conditional probabilities with probability $p^\eta_j$ (resp., $p^v_j$). The EM algorithm supplies a reliable mechanism for computing GMMs from data with guarantees on the convergence of these models \cite{Xu_GMM_convergence}. See Fig. \ref{fig:GMM2} for an example of a GMM.

The challenge, now, for anomaly detection is to accurately characterize the distribution of the residual and distance measure, ideally maintaining an analytic relationship so that the role of different design variables (e.g., observer gain) can be understood. This paper employs a generalized version of the chi-squared detector and in using the GMM representation of the noise distributions allows us to develop the GMM representation of the residual distribution and then compute statistics about the distance measure distribution. 

\subsection{Residual Distribution} 
We first characterize the residual distribution.
\vspace{2mm}
\begin{lemma} \label{lem:rk}
Given the LTI system \eqref{eq:LTI} driven by system and sensor noises whose probability density functions can be expressed as the Gaussian mixture models in \eqref{eq:noiseGMM_eta} and \eqref{eq:noiseGMM_v},
the probability density function of the residual $r_k$ at time $k\in\mathbb{N}$ can be written as a Gaussian mixture model of $m_k=m_1^km_2^{k-1}$ Gaussian modes
\begin{equation}\label{eq:kresidual}
    f_{r_k}(x)=\sum_{j=1}^{m_k} \tau_j\  \mathcal{N}(x\, |\, \beta_j,\Theta_j),
\end{equation}
where $\tau_j$ represents the mixture probabilities of the Gaussian modes, $\beta_j$ are the means, and $\Theta_j$ are the covariances,
\begin{itemize}
    \item[] $\displaystyle \beta_j = \sum_{\kappa=1}^k A_\kappa^T \mu_{n_\kappa^j}^{\eta}+\sum_{\kappa=1}^{k-1} B_\kappa^T \mu_{n_{\kappa+k}^j}^{v}$
    \vspace{2mm}
    \item[] $\displaystyle \Theta_j =\sum_{\kappa=1}^k A_\kappa K_{n_\kappa^j}^{\eta}A_\kappa^T+\sum_{\kappa=1}^{k-1} B_\kappa K_{n_{\kappa+k}^j}^{v}B_\kappa^T$
    \vspace{2mm}
    \item[] $\displaystyle \tau_j =\prod_{\kappa=1}^{k}p_{n_\kappa^j}^{\eta} \times \prod_{\kappa=1}^{k-1}p_{n_{\kappa+k}^j}^{v}$
    \vspace{3mm}
    \item[] $\displaystyle A_{\kappa}=\left\{\begin{array}{ll}I &\kappa=1\\-C(F-LC)^{\kappa-2}L &2\leq \kappa\leq k \end{array}\right.$
    \vspace{3mm}
    \item[] $\displaystyle B_{\kappa}=C(F-LC)^{\kappa-1} \qquad \,\,\, 1\leq \kappa \leq k-1$
    \vspace{1mm}
\end{itemize}
where $n_\kappa^j$ captures all the possible permutations of combining terms from the system and measurement noise GMM modes, following the rules:
\begin{itemize}[leftmargin=1.3cm]
    \vspace{1mm}
    \item[\textit{Init:}] $n_\kappa^1=1$ for $\kappa=1,\dots,2k-1$
    \vspace{3mm}
    \item[\textit{Add:}] $n_1^{j+1}=n_1^{j}+1$
    \vspace{3mm}
    \item[\textit{Wrap:}] if $n_\kappa^{j+1} > \left\{\begin{array}{lll} m_1 & \text{for} & \kappa=1,2,\dots,k \\ m_2 & \text{for} & \kappa=k+1,\dots,2k-1 \end{array}\right\}$
    \vspace{2mm}
    \item[]then $n_{\kappa+1}^{j+1}=n_{\kappa+1}^{j+1}+1$ and $n_\kappa^{j+1}=1$.
\end{itemize}
\end{lemma}
\vspace{2mm}
\begin{remark}
Since any probability density function can be expressed as a Gaussian mixture model, the key insight from Lemma \ref{lem:rk} is not that the residual can be expressed as a GMM, but rather that the residual GMM can be found analytically from the GMMs of the system and measurement noises. The notation above is unavoidably complicated due to a large number of terms. Their definitions are straightforward by looking at the proof below - namely the expansion of the product in \eqref{eq:rkfourier}. In Section \ref{application} we will discuss an effective means of reducing the number of GMM terms needed to represent the density function.
\end{remark}
\vspace{1mm} 
\begin{proof} 
In the absence of anomalies, from the definition of the residual \eqref{eq:residual_def} and the system equations \eqref{eq:LTI}, we can recursively solve for the expression of $r_k$ as a function of only the system and sensor noises (alternatively use the z-transform), 
\begin{equation} \label{eq:residual} 
r_k=\eta_k-\sum_{\kappa=1}^{k-1}C(F-LC)^{\kappa-1}(L\eta_{k-\kappa}-v_{k-\kappa}). 
\end{equation} 
Thus the residual is a linear combination of sequential iid samples of noise such that its distribution can be expressed as a convolution integral in terms of the constituent noise distributions \cite{Probability}, i.e., 
\begin{equation} \label{eq:convolution} 
f_{r_k}=f_{A_k\eta_1}*\cdots * f_{A_1\eta_k}*f_{B_{k-1}v_1}*\cdots * f_{B_1v_{k-1}} 
\end{equation} 
where $A_\kappa=-C(F-LC)^{\kappa-2}L, \quad 2 \leq \kappa \leq k$ , $A_1=I$ and $B_\kappa=C(F-LC)^{\kappa-1}, \quad 1 \leq \kappa \leq k-1$. Since the noise samples are iid, we can drop the time dependence to yield, 
\begin{equation} \label{eq:convolution2} 
f_{r_k}=f_{A_1\eta}*\cdots * f_{A_k\eta}*f_{B_1v}*\cdots * f_{B_{k-1}v}. 
\end{equation} 
The characteristic function of a random variable $X$ is defined $\varphi_X(\omega) = \mathbb{E}[e^{i\omega^T X}]$, where $\mathbb{E}[\cdot]$ denotes expectation. Using the characteristic function to represent the distributions involved in \eqref{eq:convolution2}, changes the convolution from an integral to a product. The characteristic functions of the residual and noises are, 
\begin{align} 
\varphi_{r_k}(\omega)&=\int_{-\infty}^{\infty}f_{r_k}(x)\, e^{i\omega^Tx} \ dx,\\ 
\varphi_{\eta}(\omega)&=\int_{-\infty}^{\infty}f_{\eta}(x)\, e^{i\omega^Tx} \ dx,\\ 
\varphi_{v}(\omega)&=\int_{-\infty}^{\infty}f_{v}(x)\, e^{i\omega^Tx} \ dx. 
\end{align} 
Note that the characteristic function of an affine transformation of a random variable $Y=QX+R$ is 
$$\varphi_Y(\omega)=e^{(i\omega^TR)}\varphi_X(Q^T\omega),$$ 
thus $\varphi_{A_\kappa\eta}(\omega)=\varphi_{\eta }(A_\kappa^T\omega)$ and $\varphi_{B_\kappa v} (\omega)=\varphi_{v}(B_\kappa^T\omega)$. Therefore, $\varphi_{r_k}(\omega)$ is a product of characteristic functions for transformed system and measurement noises, 
\begin{equation} \label{eq:rkfourier} 
\varphi_{r_k}(\omega)=\prod_{\kappa=1}^{k}\varphi_{\eta}\left(A_\kappa^T\omega\right) \ \times\ \prod_{\kappa=1}^{k-1}\varphi_{v}\left(B_\kappa^T\omega\right). 
\end{equation} 
Using the Gaussian mixture model with the characteristic function representation makes it possible to expand the system noise and measurement noise, e.g., for the measurement noise 
\begin{align} 
\varphi_{\eta}(\omega)&=\sum_{j=1}^{m_1} p^{\eta}_j\int_{-\infty}^{\infty}\mathcal{N}(x\, |\, \mu^{\eta}_j,K^{\eta}_j)\, e^{i\omega^Tx} \ dx, \label{eq:eta_characteristicftn1}\\ 
&= \sum_{j=1}^{m_1} p^{\eta}_j\, e^{\left(i\omega^T\mu^{\eta}_j -\frac{1}{2}\omega^T K^{\eta}_j \omega\right)}, \label{eq:eta_characteristicftn2} 
\end{align} 
where the simplification from \eqref{eq:eta_characteristicftn1} to \eqref{eq:eta_characteristicftn2} comes from identifying that the integral in \eqref{eq:eta_characteristicftn1} is the characteristic function of a Gaussian distribution which has a closed form expression \cite{Probability}. If we replace the variable $\omega$ with its transformed $A_\kappa^T\omega$ and perform the same process for the system noise, 
\begin{align} 
\varphi_{\eta}(A_\kappa^T\omega)&=\sum_{j=1}^{m_1} p^{\eta}_j\, e^{\left(i\omega^T A_\kappa \mu^{\eta}_j -\frac{1}{2}\omega^T A_\kappa K^{\eta}_j A_\kappa^T \omega\right)}, \label{eq:GMMchareta} \\ 
\varphi_{v}(B_\kappa^T\omega)&=\sum_{j=1}^{m_2} p^v_j\, e^{\left(i\omega^T B_\kappa\mu^v_j -\frac{1}{2}\omega^T B_\kappa K^v_j B_\kappa^T \omega\right)}. \label{eq:GMMcharv} 
\end{align} 
Substituting these expressions into \eqref{eq:rkfourier} reveals that $\varphi_{r_k}(\omega)$ is a linear combination of $m_k=m_1^km_2^{k-1}$ exponential terms 
\begin{equation}\label{eq:rkchar} 
\varphi_{r_k}(\omega)=\sum_{j=1}^{m_k} \tau_j\, e^{\left(i\omega^T \beta_j -\frac{1}{2}\omega^T \Theta_j \omega\right)} 
\end{equation} 
where the expressions for $\beta_j$, $\Theta_j$, and $\tau_j$ in the statement of Lemma \ref{lem:rk} can be derived by substituting \eqref{eq:GMMchareta}-\eqref{eq:GMMcharv} and expanding the product in \eqref{eq:rkfourier}. Conveniently, \eqref{eq:rkchar} is in the form of the characteristic function of a linear combination of Gaussian functions, thus \eqref{eq:rkchar} demonstrates that the residual distribution at time step $k$ can be expressed as 
\begin{equation} \label{eq:residual_convergence} 
f_{r_k}(x)=\sum_{j=1}^{m_k} \tau_j\, \mathcal{N}(x\, |\, \beta_j,\Theta_j), 
\end{equation} 
where $\tau_j$ represents the mixture probabilities of the $m_k$ Gaussian modes, $\beta_j$ are the means of the modes, and $\Theta_j$ are the covariances of the modes.
\end{proof} 
\begin{figure*}[t] 
\begin{tcolorbox}[colback=white] 
\begin{equation}\label{eq:zkcdffinal_M}\tag{$\star$} 
\begin{aligned} 
M_j&=\frac{\det{\mathcal{C}}}{\sqrt{(2\pi)^p\det{K_j}}}\int_{0}^{2\pi}\int_{0}^{\pi} \cdots \int_{0}^{\pi} \int_{0}^{\sqrt{\alpha}} e^{(\mathcal{C}^T \rho_k +\gamma_j)^T K_j^{-1}(\mathcal{C}^T \rho_k +\gamma_j)}\ d \boldsymbol{\rho}\\\vspace{2mm} 
d \boldsymbol{\rho}&=|\rho_k|^{p-1}\sin{(\phi_1)}^{p-2}\sin{(\phi_2)}^{p-3}\cdots \sin{(\phi_{p-2})}\ d|\rho_k|\ d{\phi_1}\ d{\phi_2}\ ...\ d{\phi_{p-1}} 
\end{aligned} 
\end{equation} 
\end{tcolorbox} 
\end{figure*} 
\vspace{1mm} 
In the absence of anomalies, the time dependence of the residual is governed by the convergence of the estimator, since the system is time-invariant and the noises are iid. In most practical situations the estimator is designed to converge relatively quickly and so it is reasonable to assume for the rest of this work that sufficient convergence of the estimation has already been achieved. Thus we seek the steady state distribution of the residual, which permits some simplification by removing the dependence on time. 
\vspace{1mm} 
\begin{lemma} \label{assymone} 
If the system is stable, then the distribution of the residual converges to $f_{r_\infty}$ as $k\to\infty$, i.e., given some error tolerance $\epsilon\in\mathbb{R}_{>0}$ there exists a $k^*\in\mathbb{N}$ such that 
\begin{equation} \label{eq:rk_lemma}
\|f_{r_\infty} - f_{r_{k^*}} \| < \epsilon, 
\end{equation} 
implying that $f_r := f_{r_{k^*}}$ provides an arbitrarily close approximation of the steady state distribution. 
\end{lemma} 
\vspace{1mm} 
\begin{proof} 
To not belabor an intuitive result, we provide a sketch of the proof without explicitly writing all details regarding the convergence. Because the system is stable $\rho({F-LC}) <1$, which means 
$$\lim_{\kappa \to \infty} A_\kappa = O_{p\times p}, \quad\text{and}\quad \lim_{\kappa \to \infty} B_\kappa = O_{n\times p},$$ 
where $O$ is the zero matrix of the designated size. Therefore, in steady state the characteristic functions characterizing the system and measurement noise become 
\begin{align} 
\lim\limits_{\kappa \to \infty} \varphi_{\eta}(A_\kappa^T\omega)&=\lim\limits_{\kappa \to \infty} \sum_{j=1}^{m_1} p^{\eta}_j\, e^{(i\omega^T A_\kappa\mu^{\eta}_j -\frac{1}{2}\omega^T A_\kappa K^{\eta}_j A_\kappa^T \omega)} \nonumber \\ 
&= \sum_{j=1}^{m_1} p^{\eta}_j=1,\\ 
\lim\limits_{\kappa \to \infty} \varphi_{v}(B_\kappa^T\omega)&=\lim_{\kappa \to \infty}\sum_{j=1}^{m_2} p^v_j\, e^{(i\omega^T B_\kappa\mu^v_j -\frac{1}{2}\omega^T B_\kappa K^v_j B_\kappa^T \omega)} \nonumber \\ 
&= \sum_{j=1}^{m_2} p^v_j=1. 
\end{align} 
The convergence of these characteristic functions imply the convergence of the characteristic function of the residual to $\varphi_{r_{\infty}}(\omega)$ such that for a given $\tilde{\epsilon}\in\mathbb{R}_{>0}$ there exists a $k^*$ such that 
\begin{equation} 
|\varphi_{r_\infty}(\omega) - \varphi_{r_{k^*}}(\omega)| < \tilde{\epsilon}, 
\end{equation} 
where $\varphi_{r_k}$ is defined as in \eqref{eq:rkfourier}. This provides an approximation for the steady state that is made arbitrarily accurate by selecting an arbitrarily small $\tilde{\epsilon}$ (and hence large $k^*$), 
\begin{align} 
\varphi_{r_{\infty}}(\omega)=& \prod_{\kappa=1}^{\infty}\varphi_{\eta}(A_\kappa^T\omega)\times\prod_{\kappa=1}^{\infty}\varphi_{v}(B_\kappa^T\omega)\\ 
\approx&\prod_{\kappa=1}^{k^*}\varphi_{\eta}(A_\kappa^T\omega)\times\prod_{\kappa=1}^{k^*-1}\varphi_{v}(B_\kappa^T\omega). \label{eq:approx_rkstar_charftn} 
\end{align} 
As mentioned earlier, $k^*$ can be interpreted as the settling time of the control system and estimator. As before, the characteristic function of the residual corresponds to a probability density function that is composed of the sum of multiple Gaussian modes - a GMM. This distribution can also be made arbitrarily accurate by selecting a smaller $\tilde{\epsilon}$ which corresponds to a larger $k^*$ and a smaller $\epsilon$ in \eqref{eq:rk_lemma}, 
\begin{equation} \label{eq:finalresidual} 
f_{r_{\infty}}(x)\approx f_r(x) := f_{r_{k^*}}(x) = \sum_{j=1}^m \pi_j\ \mathcal{N}(x\,|\,\mu_j,K_j), 
\end{equation} 
where $m=m_1^{k^*}m_2^{k^*-1}$ is the number of Gaussian modes used to represent the steady state residual distribution. As before, the values of $\pi_j$, $\mu_j$, and $K_j$ are the mixture probabilities, means, and covariances, respectively, of the $m$ Gaussian modes and are computed by substituting \eqref{eq:GMMchareta}-\eqref{eq:GMMcharv} and expanding the product in \eqref{eq:approx_rkstar_charftn}. 
\end{proof} 

\begin{figure*}[t]
\begin{center} 
\includegraphics[width=\linewidth]{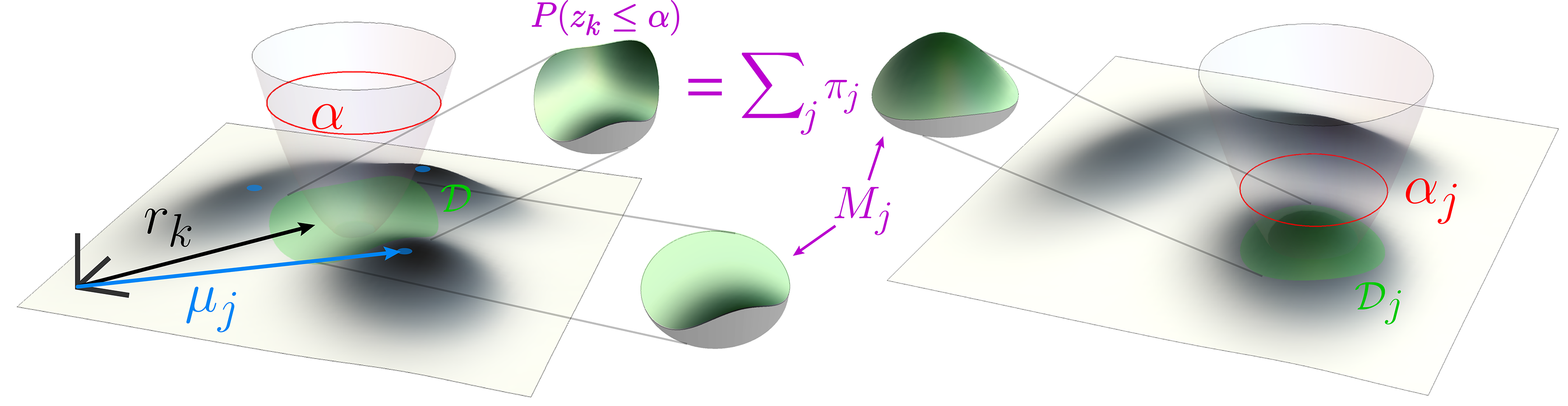} 
\end{center} 
\caption{The 3D surface of the residual GMM probability density function (shaded to reveal height mapping) constructed from $m=3$ Gaussian modes with means $\mu_j$ denoted by blue dots. The quadratic form of the distance measure $z_k$ forms a paraboloid over the domain of the residual $r_k$ centered at $\mu$ (the mean of the residual distribution). The cumulative probability $P(z_k\leq \alpha)$ corresponds to the amount of probability in the residual distribution within the region $\mathcal{D}$ \eqref{eq:D} (green area), the area corresponding to the domain formed by projecting the $z_k=\alpha$ level set of the paraboloid (red line). The probability $P(z_k\leq \alpha)=\sum_j \pi_jM_j$, where $M_j$ is the volume of probability contributed by the Gaussian mode $j$. In Section \ref{relation}, we show that we can interpret the generalized detector as the probabilistic combination of $m$ chi-squared detectors, where the $j$-th detector is tuned such that the false alarm rate is $\mathcal{A}_j=1-M_j$, which means the threshold $\alpha_j$ is selected such that the volume of probability contained within the region $\mathcal{D}_j$ \eqref{eq:Dj} (the projection of the $\alpha_j$-level set of the paraboloid centered at $\mu_j$) is equal to $M_j$.} \label{fig:zkrk} 
\end{figure*} 

\subsection{Distance Measure \& False Alarm Rate Tuning}\label{distance} 
Most detectors construct a scalar-valued distance measure from the residual to quantify how different the measurement is from what is expected. In this paper we use the generalized chi-squared detector in \eqref{eq:chisquared} where $\mu$ is the overall mean value and $\Sigma$ is the overall covariance of the steady state residual. From \eqref{eq:finalresidual}, 
\begin{equation} 
\mu=\sum_{j=1}^m \pi_j \mu_j \quad\text{and}\quad 
\Sigma=\sum_{j=1}^m \pi_jK_j+\gamma_j \gamma_j^T 
\end{equation} 
where $\gamma_j=\mu-\mu_j$ is the difference between the individual GMM (mode) means and overall mean. Our aim in this section, and ultimately of this paper, is to characterize the expected rate of false alarms given a chosen threshold value, $\alpha$, of the detector. Recall that alarms are generated if $z_k>\alpha$ for any $k\in\mathbb{N}$. Thus the probability of drawing a distance measure value from its distribution that leads to an alarm is $P(z_k > \alpha)$. \figref{fig:zkrk} plots the multivariate, scalar-valued distance measure function $z_k$ over the domain of the residual $r_k$, in the two dimensional ($p=2$) case. Due to the quadratic form of the distance measure in \eqref{eq:chisquared}, the surface is a paraboloid. The region $\mathcal{D}$ is the area contained by the projection of the level set $z_k=\alpha$ onto the $r_k$ plane. Theorem \ref{lem:generalizedlem1} is the generalized version of Lemma \ref{lem:chisquared_tuning} for tuning the detector for a desired level of performance (desired false alarm rate) in the case that the system and measurement noises are no longer Gaussian. 

\vspace{1mm} 
\begin{theorem} \label{lem:generalizedlem1} 
Assume that there are no anomalies present in an LTI system \eqref{eq:LTI} driven by arbitrary system and measurement noises such that the residual $r_k \sim f_r$ \eqref{eq:finalresidual}, i.e., an $m$-th order Gaussian mixture model, and consider the generalized chi-squared detector \eqref{eq:chisquared} with threshold $\alpha \in \mathbb{R}_{>0}$. The expected false alarm rate is
\begin{equation} \label{eq:zkcdffinal} 
\mathcal{A}=1-P(z_k \leq \alpha) =1-\sum_{j=1}^m \pi_j M_j, 
\end{equation} 
where $M_j$ is given by \eqref{eq:zkcdffinal_M}, in the box above. 
\end{theorem} 
\vspace{1mm} 
\begin{proof} 
The false alarm rate is the probability that $z_k$ exceeds the threshold $\alpha$, $\mathcal{A}=P(z_k>\alpha)=1-P(z_k \leq \alpha)$. The cumulative probability of the nonnegative distance measure is given by 
\begin{equation}\label{eq:cumu} 
P(z_k \leq \alpha) = \int_0^\alpha f_{z_k}(z)\ dz = \iint_\mathcal{D} f_{r}(\boldsymbol{r}) \ d \boldsymbol{r},
\end{equation} 
where $d \boldsymbol{r}$ is the differential area element over the region 
\begin{equation} \label{eq:D}
\mathcal{D}=\{r_k\,|\, z_k=(r_k-\mu)^T\Sigma^{-1}(r_k-\mu)\leq \alpha\},
\end{equation}
which is in general a $p$-dimensional ellipsoid and whose boundary is defined by the projection of the level set $z_k=\alpha$ onto the $r_k$ plane, and threshold $\alpha$ is the assigned threshold of the generalized chi-squared detector. Effectively, \eqref{eq:cumu} expresses that the probability of having $z_k\leq\alpha$ is equal to the volume under the $f_r$ distribution, restricted to the $r_k$ values that generate a $z_k$ value less than or equal to $\alpha$. Since $f_r$ is composed of a mixture of Gaussian modes, Fig. \ref{fig:zkrk} depicts how this formulation sums the volume of probability contained under the various Gaussian modes and within the region $\mathcal{D}$. Replacing $f_{r}$ in equation \eqref{eq:cumu} explicitly expressing the $m$ GMM modes leads to 
\begin{equation*} 
\begin{aligned} 
&P(z_k \leq \alpha) =\\ 
&\frac{1}{\sqrt{(2\pi)^p}}\sum_{j=1}^m \frac{\pi_j}{\sqrt{\det{K_j}}}\iint_{\mathcal{D}} e^{(r_k-\mu_j)^T K_j^{-1}(r_k-\mu_j)} d \boldsymbol{r}. 
\end{aligned} 
\end{equation*} 
In order to write this equation in terms of the assigned threshold on the detector, we write it in normalized spherical form, changing the volume element $d \boldsymbol{r}$ to the volume element $d\boldsymbol{\rho}$ which is a volume element over an $p$-sphere characterized with radius $\sqrt{\alpha}$; hence, 
$$r_k-\mu=\mathcal{C}^T\rho_k \quad \to \quad r_k-\mu_j= \mathcal{C}^T \rho_k+\gamma_j,$$ 
$$\text{and}\quad d \boldsymbol{\rho}=\frac{d \boldsymbol{r}}{\det{\mathcal{C}}},$$ 
where $\rho_k$ is a vector varying inside the $p$-sphere and is an affine transformation of the steady state residual $\rho_k=\mathcal{C}^{-T}(r_k-\mu)$ and the matrix $\mathcal{C}$ is the Cholesky decomposition of covariance $\Sigma$. 
This transformation simplifies the new region of integration to be a $p$-sphere. Making this substitution and expressing the limits of integration in spherical form yield the final from in \eqref{eq:zkcdffinal} and \eqref{eq:zkcdffinal_M}, where $\rho_k$ in spherical form is a function of its norm $|\rho_k|$ and angles $\phi_i$, $i=1\dots p-1$, where $p$ is the dimension of the measurement.
\end{proof} 
\vspace{1mm} 
\begin{remark} 
Now that we have the relationship \eqref{eq:zkcdffinal} and \eqref{eq:zkcdffinal_M} between assigned threshold of the detector $\alpha$ and the corresponding CDF for the distance measure in the no fault/attack case $P(z_k<\alpha)$, we can use this result to find the threshold $\alpha$ that provides a desired false alarm rate. There is a mapping between the threshold $\alpha$ and the false alarm rate $\mathcal{A}$. Armed with Theorem \ref{lem:generalizedlem1}, it is possible to use, for example, a bisection method to find the threshold value to yield a desired false alarm rate. 
\end{remark} 

\section{INTERPRETATION} \label{relation} 
Using an approach that leverages the Gaussian mixture model representation of arbitrary noise distributions not only recovers analytic tractability, it also provides an intuitive and instructive backwards compatibility with the standard chi-square detector driven by Gaussian noise. Suppose that in using the tools presented in this paper, we find that the GMM of the residual distribution is the combination of three distinct Gaussian modes (e.g., see Fig. \ref{fig:GMM2}), 
\begin{equation}\label{eq:3residue} 
f_{r}(x)=\sum_{j=1}^{3} \pi_j\, \mathcal{N}(x\,|\,\mu_j, K_j), 
\end{equation} 
where $\mu_j$ are the mean values of the GMM modes, $K_j$ are the corresponding covariances, and $\pi_j$ are the mixing probabilities. 

A way to interpret the GMM residual probability distribution \eqref{eq:3residue} is that at each time $k$ the residual $r_k$ is drawn from the first GMM mode with probability $\pi_1$, drawn from the second GMM mode with probability $\pi_2$, and drawn from the third GMM mode with probability $\pi_3$. For each of these modes of the residual GMM, we can reverse engineer a hypothetical Gaussian measurement noise that if applied to the system in isolation would generate a Gaussian residual equal to that mode of the residual GMM. This hypothetical Gaussian noise would have mean $a_j$ and covariance $C_j$, 
\begin{equation}\label{eq:designednoise} 
a_j=E^{-T}\mu_j \quad\text{and}\quad C_j=E^{-1}K_jE^{-T}, 
\end{equation} 
for $j\in\{1,2,3\}$ and the matrix $E$ is defined as 
$$E=\sum_{\kappa=1}^{\infty} A_\kappa.$$ 
If the measurement noise was Gaussian with mean $a_j$ and covariance $C_j$, $j\in\{1,2,3\}$, then we could tune a conventional chi-squared detector using Lemma \ref{lem:chisquared_tuning} to determine a threshold $\alpha_j$ to yield a false alarm rate $\mathcal{A}_j$. If further, we selected $\mathcal{A}_j=1-M_j$, where $M_j$ is defined in \eqref{eq:zkcdffinal_M}, then \eqref{eq:zkcdffinal} becomes
\begin{align}\label{eq:imaginarycdf} 
\mathcal{A} &= 1 -\pi_1(1-\mathcal{A}_1) - \pi_2(1-\mathcal{A}_2) - \pi_3(1-\mathcal{A}_3),\\
&=\pi_1\mathcal{A}_1 + \pi_2\mathcal{A}_2 + \pi_3\mathcal{A}_3, 
\end{align}
since $\pi_1+\pi_2+\pi_3=1$. In this context, then $M_j$ can also be interpreted the probability under the $j$-th Gaussian mode distribution (characterized by $\mu_j$ and $K_j$) over the integration region
\begin{equation} \label{eq:Dj}
    \mathcal{D}_j := \{r_k\,|\, z_k=(r_k-\mu_j)^T K_j^{-1}(r_k-\mu_j)\leq \alpha_j\},
\end{equation} 
which is defined by the level set $z_k=\alpha_j$ of the paraboloid characterized by $\mu_j$ and $K_j$. Thus $M_j$ and $\alpha_j$ are related according to Lemma \ref{lem:chisquared_tuning} 
\begin{equation}\label{eq:alphai} 
\alpha_j=2P^{-1}\left(\frac{M_j}{2},\frac{p}{2}\right). 
\end{equation}
This interpretation is depicted in Fig. \ref{fig:zkrk}.

\section{EXAMPLE} \label{application} 
\begin{figure}[t] 
\centering 
\includegraphics[width=0.9\linewidth]{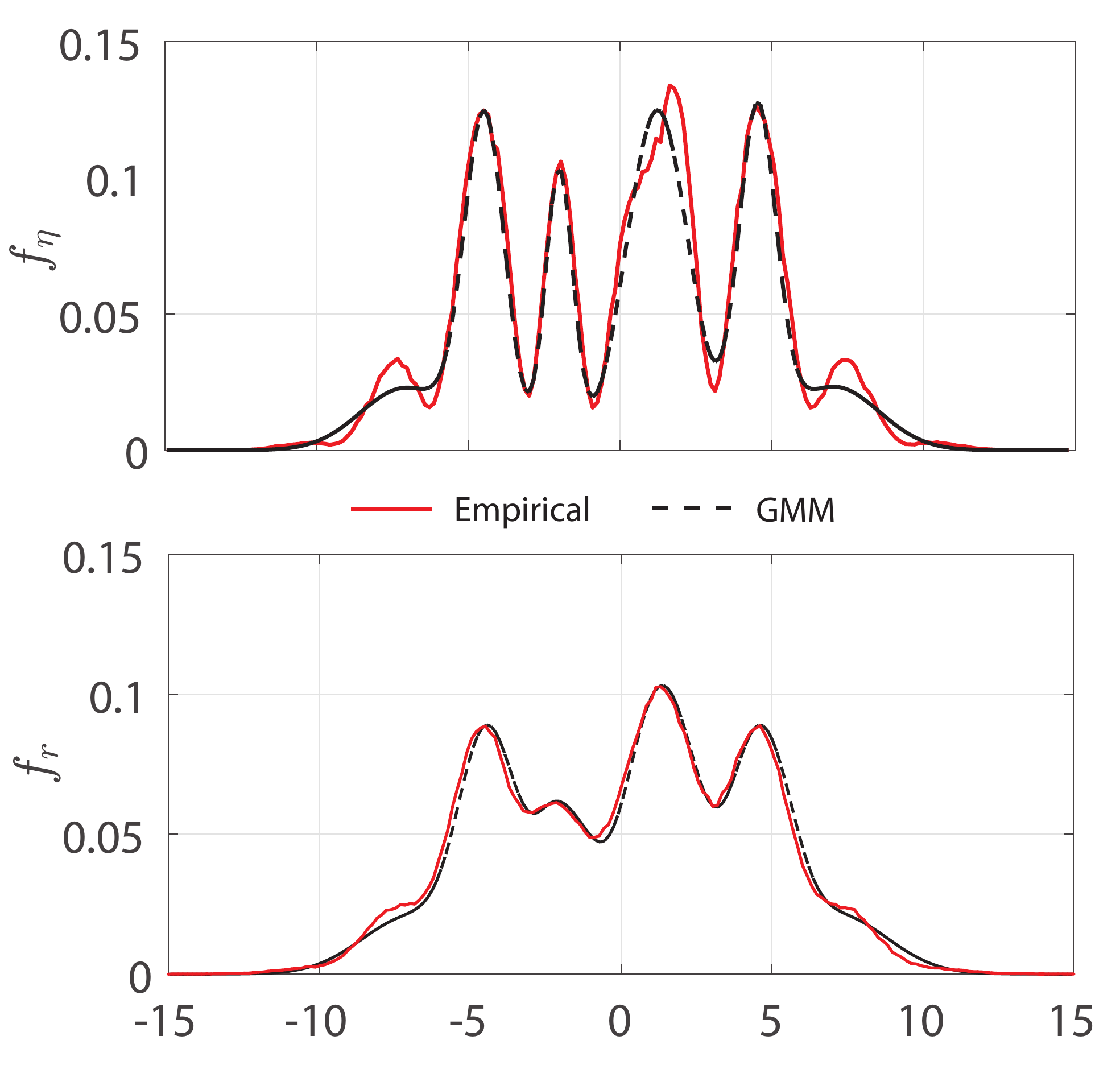}
\caption{The measurement noise distribution (top) is approximated by a Gaussian mixture model with six modes and leads to a corresponding complex residual distribution (bottom). The GMM approximations agree well with the empirical distributions from Monte-Carlo simulations.} 
\label{fig:example_system}
\end{figure}

Consider a single output system and estimator characterized by the following matrices and driven by measurement noise (no system noise) distributed according to the probability density function shown in Fig. \ref{fig:example_system},
\begin{equation*} \label{eq:example_system} 
F= \begin{bmatrix} 
0.8 & 0.2 \\ 
-0.25 & 0.1 
\end{bmatrix},\ \ 
C= \begin{bmatrix} 
0.5 & 0.5 
\end{bmatrix}, \ \ 
L= \begin{bmatrix} 
0.3\\ 
-0.3 
\end{bmatrix}. 
\end{equation*} 
We select a threshold for the detector $\alpha = 0.75$ and use Theorem \ref{lem:generalizedlem1} to calculate the expected false alarm rate for this threshold. 

Figure \ref{fig:example_system} shows the fit of the GMM of the measurement noise compared to the empirical distribution attained from a Monte-Carlo simulation with $5\times10^6$ samples. Here we select a mixture of six ($m_1=6$) Gaussian modes and Table \ref{tab:GMMvalues} presents the means and covariances of each mode. 

\begin{figure}[t] 
\begin{center} 
\includegraphics[width=1\linewidth]{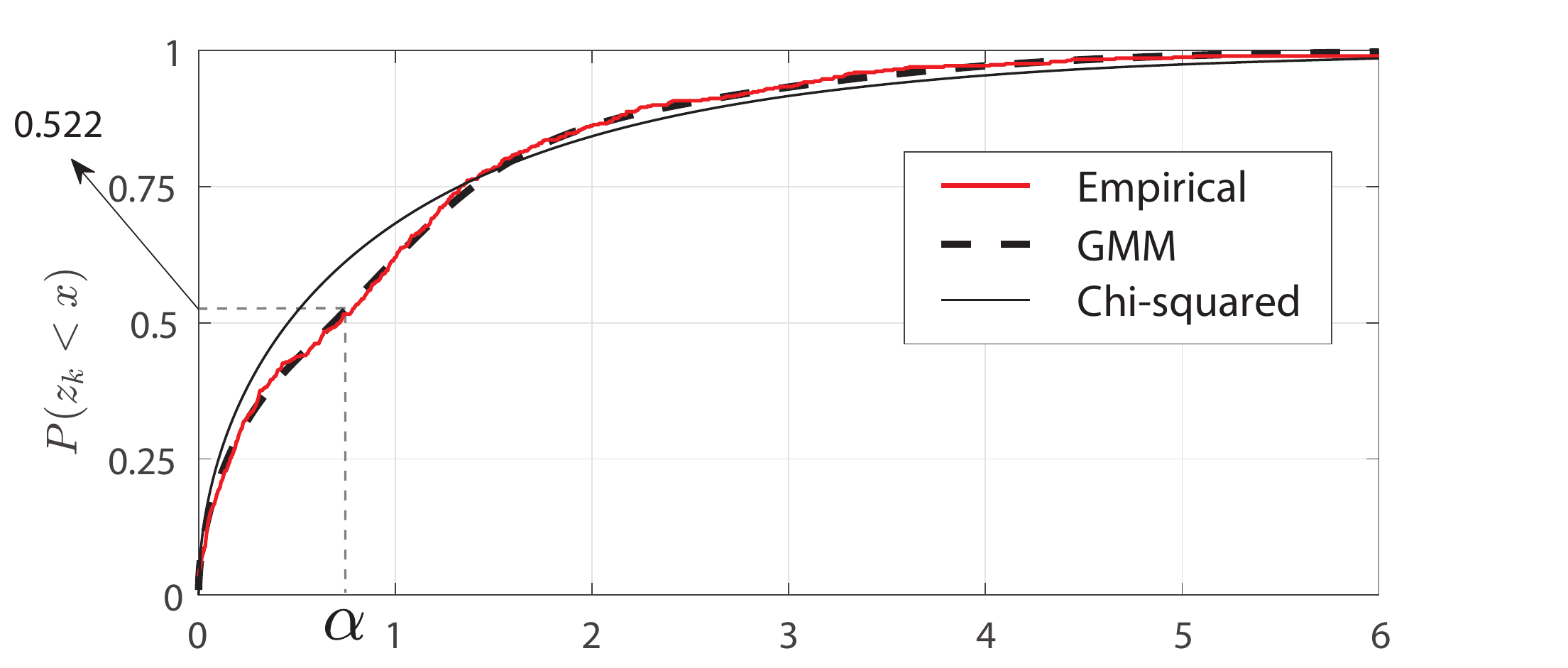} 
\end{center} 
\caption{The cumulative distribution function of the distance measure corresponding to the measurement noise in Fig. \ref{fig:example_system} evidencing noticeable deviations away from a comparable chi-squared distribution (which would correspond to Gaussian measurement noise). The GMM distribution agrees very well with the empirical distribution found through Monte-Carlo simulation. For a threshold $\alpha=0.75$, the GMM distribution, using Theorem \ref{lem:generalizedlem1}, predicts a false alarm rate of $\mathcal{A}= 1-0.522=0.478$. The empirical distribution has a false alarm rate of $\mathcal{A}= 1-0.516=0.484$.} 
\label{fig:cdf} 
\end{figure}

One of the challenges of our method is the possibility of having a large number of terms in the residual distribution GMM: selecting the settling time $k^*=10$, $m=m_1^{k^*}=6^{10}$ since we have no system noise. In practice, however, many of these terms have mean and covariance values that are extremely similar. We can then greatly simplify the GMM expression by aggregating terms that are roughly the same. To do this we define a threshold on the normed difference between mean values and covariance values, $d_{\mu}$ and $d_K$.
If the normed difference between mean values and covariance values for any pair of modes is less than these thresholds, 
$$\|\mu_i-\mu_j\|\leq d_\mu\qquad\text{and}\qquad \|K_i-K_j\|\leq d_K,$$
for $i\neq j\in\{1,\dots,m_1\}$, we consider those terms the same and their coefficients are added together. This procedure provides an arbitrarily accurate approximation with significantly fewer terms. Here we select $d_\mu=0.0747$ and $d_K=0.0917$ using a heuristic based on the spread of the distribution and these choices lead to $\widetilde{m}=282$ terms that remain, significantly fewer than the original $m =6^{10}$.
\begin{table}[t] 
\begin{tabular}{|l|c|c|c|c|c|c|} 
\hline 
& 1& 2& 3& 4& 5& 6 \\ \hline 
$p^{\eta}$& 0.0847 & 0.2012 & 0.1184 & 0.3200 & 0.1889 & 0.0869 \\ \hline 
$\mu^{\eta}$&-7.0877 & -4.4709 & -2.0082 & 1.2318 & 4.5240 & 7.0504 \\ \hline 
$K^{\eta}$&2.1997 &0.4471 &0.2062 &1.0392 & 0.3858 &2.2329 \\ \hline 
\end{tabular} 
\caption{The coefficients, means, and covariances of the GMM modes ($m_1=6$) of the measurement noise in Fig. \ref{fig:example_system}.} \label{tab:GMMvalues} 
\end{table} 
Using this reduced number of modes, the GMM expression of the distribution of residual is shown in Fig. \ref{fig:example_system} and compares favorably with the empirical residual distribution computed (through Monte-Carlo simulation) from the true noise distribution. By using Theorem \ref{lem:generalizedlem1}, we calculate the expected false alarm rate
\begin{equation*} 
\mathcal{A}=1-P(z_k\leq 0.75)=0.478 
\end{equation*} 
This expected false-alarm rate is shown in Fig. \ref{fig:cdf} and the value of $\mathcal{A}$ compares well with the value attained through Monte-Carlo simulation ($0.484$).
\section{CONCLUSION} \label{results} 
In this paper, for discrete-time LTI systems subject to arbitrary sensor and measurement noise, we provide tools to tune model-based detectors and characterize the trade-off between true and false positives. We have generalized one of the most widely used fault detector, the chi-squared detector, for use with general noise distributions. Our approach uses a Gaussian mixture model expression of the disturbances which preserves some of the appealing analytic tractability of working with Gaussian noises on an LTI system. 
\bibliographystyle{IEEEtran} 
\bibliography{security} 

\end{document}